\documentstyle[preprint,seceq]{jpsj}
\newcommand{\tone}{ spin-$\frac{1}{2}$-$\frac{1}{2}$-1-1 }
\newcommand{\hyp}{ \rule[0.6mm]{1mm}{0.1mm} }
\newcommand{\simle}{ \: \raisebox{0.1cm}{\tiny $<$} \!\!\!
\raisebox{-0.05cm}{\tiny $\sim$} \:}

\title{
Ground State Property of an Alternating\\
Spin Ladder Involving Two Kinds of Inter-Chain Interactions}

\author{
Atsuo {\sc Satou}\thanks{e-mail: atsuo@grad.ap.kagu.sut.ac.jp}
and Yoshiko Oi {\sc Nakamura}}

\inst{
Department of Applied Physics, Faculty of Science,
Science University of Tokyo,\\
1-3 Kagurazaka, Shinjuku-ku, Tokyo 162-8601 }

\recdate{\today}

\abst{
The ground state property of the alternating spin ladder is studied in the case that the
system involves an antiferromagnetic intra-chain interaction as well as two kinds of
inter-chain interactions; one is between spins of the same magnitude and the other is between
spins with different magnitudes. The calculation has been carried out by the exact
diagonalization method. As a consequence of the competition among interactions, the system
is revealed to show an interesting variety of phases in the ground state property. Its
phase diagram is exhibited in the parameter space of the system. We find that, however
small the total amount of the inter-chain interactions is, the ferrimagnetic ground state
becomes unstable in a certain region. In this case, which of the ferrimagnetic and the
singlet ground state to appear is determined only by the ratio between the inter-chain interactions
regardless of their total amount. The nature of two phases appearing in the singlet region
of the phase diagram and the type of the phase transition between them are also discussed.
The results are ensured by comparing with those of obtained in other models which are
contained in our model as special limiting cases.}

\kword{
alternating spin chain, alternating spin ladder, quantum ferrimagnet, quantum mixed \tone
chain, exact diagonalization, quantum phase transition}

\def\runtitle{Ground State Property of an Alternating
Spin Ladder Involving Two Kinds of Inter-Chain Interactions}

\def\runauthor{ Atsuo {\sc Satou} and Yoshiko Oi {\sc Nakamura} }

\begin{document}
\sloppy
\maketitle

\section{Introduction}

The prediction given by Haldane in his distinguished works \cite{haldane1,haldane2} has had
a profound influence on the research in the low-dimensional spin systems during the
following decades. In the subsequent decade, extensive studies on this issue had
been developed by means of various theoretical approaches as well as  computational
ones. As a result, the prediction seems now to be entirely agreed with and has
received wide recognition \cite{nenp,gapmonte}. This prediction states that in the
one-dimensional antiferromagnetic Heisenberg model, qualitative properties of the
ground state depend entirely on whether the system is constructed of half-integral
spins or integral spins. The energy gap, named Haldane gap, does exist between the
ground state and the first excited state in the case of integral spin, while the
energy gap never appears in the case of half-integral spin. During recent years in
this decade, the current interest on this issue has extended to a various kinds of
low-dimensional spin systems; mixed spin chains with various arrangements of spins
and bonds as well as spin-ladder systems of various kind. This renewal of the
interest on this matter owes partly to the advance of experimental techniques and
the recent technical development in the material science; the discoveries and
experimental realizations of such various kinds of low-dimensional systems in
recent years have attracted widespread attention beyond its own research field and
have inspired accumulating efforts and activities in this field.

Among these low-dimensional spin systems, an alternating spin chain with
antiferromagnetic nearest neighbor interactions have provided rich subjects in the
field from both theoretical and experimental aspects.
\cite{matrix,low-temp,pati1,pati2,mag1,ground,mag2,elementary,thermo,combi,low-ene,
mag3,bimetallic1,bimetallic2,mag4}
The spin arrangement of this alternating spin chain
is such that the constituent spins of different magnitude align alternately along
the chain. Because of this spin arrangement and the antiferromagnetic nearest
neighbor interaction, the application of the Lieb-Mattis theorem \cite{lieb}
leads to that the total spin moment of this chain has a non-zero value in its
ground state; so-called ``quantum ferrimagnetic state'' is realized as the ground
state. It is also shown by the spin wave theory that this alternating spin chain
has two kinds of excitations, one is ferromagnetic and the other antiferromagnetic;
because of these excitations thermodynamic properties of the alternating spin chain
shows ferromagnetic-antiferromagnetic crossover \cite{thermo,combi,low-ene,mag3}.

From experimental side, on the other hand, several kinds of the alternating spin chains
have been found in actual materials recently \cite{bimetallic1,bimetallic2,mag4}, and
detailed studies on their lattice structures and their magnetic properties have
been developed extensively. For example, the compound, MnCu-pbaOH that is
MnCu(pbaOH)(H$_2$O)$_3$
with pbaOH $=$ 2-hydroxy-1,3-propylenebis (oxamato), has such crystal
structure that Mn$^{2+}$ ($S= \frac{5}{2}$) and Cu$^{2+}$ ($S= \frac{1}{2}$) ions
bridged by oxamato groups are forming antiferromagnetic alternating spin chain along
the {\it b}-axis, named by Mn$^{\rm II}$Cu$^{\rm II}$ bimetallic chain
\cite{bimetallic1}. Comparing with the intra-chain interaction,
the inter-chain exchange interactions between the nearest spins (Mn-Mn and
Cu-Cu) seems to be rather weak. Therefore we may regard this compound as an assembly
of $\frac{1}{2}$-$\frac{5}{2}$-spin alternating chains which are interacting weakly
each other so that we can expect for the ground state of this compound to be
ferromagnetic. In fact, the results of magnetic measurements on this compound are
consistent with this view \cite{bimetallic1}. So we may list this
Mn$^{\rm II}$Cu$^{\rm II}$ bimetallic chain as an example of the actual realization
of a ``quantum ferrimagnet''. It should be noted, however, that this kind of
antiferromagnetic bimetallic chains realized in actual compounds does not always
exhibit the ferrimagnetic character. There exist, in fact, another family compound
which contains the same Mn$^{\rm II}$Cu$^{\rm II}$ bimetallic chains, but it has
nevertheless a non-magnetic ground state; MnCu-pba that is
MnCu(pba)(H$_2$O)$_3 \cdot 2$H$_2$O
with pba $=$ 1,3-propylenebis (oxamato) is just the case \cite{bimetallic1}.
There is not a great difference between crystal structure of MnCu-pbaOH and MnCu-pba
except for relative positions of the chains along the {\it a}-axis. In the MnCu-pba
as compared with the MnCu-pbaOH, every chains make a slight shift along the
{\it b}-axis ( the chain direction ) relatively to their respective neighboring
chains. As a consequence of this shift, the inter-chain interaction between Mn-Cu in
the MnCu-pba may become weak compared with that in the MnCu-pbaOH, and the inter-chain
interaction between Mn-Mn and Cu-Cu may become strong at the same time. These
considerations, therefore, lead us to an idea that the competition between two kinds of
inter-chain interactions play an important roll in the determination of magnetic
properties of these compounds at low temperatures.

The aim of this paper is to clarify the magnetic character of such materials
containing spin alternating chains as stated above. Actually in this paper, for the
sake of simplicity, we investigate the ground state properties of a two-leg ladder
system composed of alternating spin chains, called by alternating spin ladder. There,
we introduce two kinds of inter-chain interactions, one is between same kind of spin
and the other between different kinds of spins. Then we examine the effect of the
competition among the interactions on the properties of the ground states, and make
up its phase diagram by the use of numerical method, the exact diagonalization using
the Lanzos method.

In the next section, we present the model Hamiltonian and explain some of results
derived from it by simple considerations before we undergo the numerical analysis.
The discussion is also made comparing our model with other ones studied by others
previously. In \S 3, are shown the results derived numerically with this model
Hamiltonian. In \S 4, we exhibit the resultant phase diagram for the ground
state of the alternating spin ladder and discuss what it means. Finally, the
results are summarized in \S 5.

\section{The Model}

Here we present the model Hamiltonian for a spin ladder composed of the same kind of
alternating spin chains which are interacting by two kinds of inter-chain interactions.
For the sake of simplicity, the magnitudes of spins constructing the constituent
alternating spin chain are taken actually as $S = \frac{1}{2}$ and $1$, since the
properties of the alternating spin chain is qualitatively independent of the magnitudes
of spins.\cite{pati2} The Hamiltonian is given as follows;
\begin{eqnarray}
H &=& J \sum_{j=1}^2 \sum_{i=1}^{\frac{N}{2}}
\mbox{\boldmath $S_{j,i}$}
\cdot \mbox{\boldmath $S_{j,i+1}$}
+ J_0(1- \delta) \sum_{i=1}^{\frac{N}{2}}
\mbox{\boldmath $S_{1,i}$} \cdot \mbox{\boldmath
$S_{2,i}$} \nonumber\\
  &+& J_0(1+ \delta) \sum_{i=1}^{\frac{N}{2}}
\mbox{\boldmath $S_{1,i}$} \cdot
\mbox{\boldmath $S_{2,i+1}$},
\end{eqnarray}
where index j specifies the chain number and $\mbox{\boldmath $S_{j,2i-1}$}$ and
$\mbox{\boldmath $S_{j,2i}$}$ denote spins of magnitude $\frac{1}{2}$ and $1$ on each
chain, respectively. Here $N$ is the total number of spins in this system, and is
taken as a multiple of four because of a periodic boundary condition
( $\mbox{\boldmath $S_{j, \frac{N}{2}+1}$} = \mbox{\boldmath $S_{j,1}$}$ ) imposed in
this paper. $J$ represents the intra-chain interaction in each alternating spin chain.
$J_0(1- \delta)$ and $J_0(1+ \delta)$ are the inter-chain interactions between the
chains; the former represents interaction between the same kind of spins ($1$-$1$ and
$\frac{1}{2}$-$\frac{1}{2}$) and the latter between different kinds of spins
($1$-$\frac{1}{2}$), respectively. A schematic representation of the Hamiltonian is
given in Fig.1(a). Since we are to investigate the effect of the competition among
the antiferromagnetic interactions on the ground state properties, we restrict the
parameters $J_0$ , $J$ and $\delta$ to the following region; $J_0 > 0$, $J > 0$ and
$ 1 \ge \delta \ge -1$. Here we should note that changing the value of $\delta$
corresponds to the relative shift between chains along the chain direction.

Even though we restrict the parameter region as stated above, the Hamiltonian still
maps to various models if we change the parameters. In our parameter region, there
exists such a region that we can apply there the Lieb-Mattis theorem to the system
and are able to find the total spin momentum of the ground state. For each case of
$\delta = 1$ and $-1$, the Hamiltonian contains only one kind of inter-chain
interaction and is free from the frustration so that we can use the Lieb-Mattis
theorem to investigate the ground state properties: As a result, we find that the
ground state is a singlet state when $\delta = -1$ and, on the other hand, the
ground state becomes a quantum ferrimagnetic state having a total spin moment
($= \frac{N}{4}$) when $\delta = 1$. From these results for two limiting cases
($\delta = 1$ and $-1$), therefore, it is easy to understood that the phase transition
between the singlet state and the quantum ferrimagnetic state should occur between
$\delta = 1$ and $\delta = -1$ for any value of the ratio $J_0/J$.

Further, we can show another exact result about the ground state of the alternating
spin ladder under a certain constraint on the parameters; when $J_0(1+ \delta) = 2J$
with $J_0 \ge 3J$, the exact ground state is a dimer state which is depicted in
Fig.2(a) by means of the valence-bond-solid picture. The proof can be made almost in
the same way as Shastry and Sutherland have done \cite{shastry}: In the case of
$J_0(1+ \delta ) = 2J$, the Hamiltonian of the alternating spin ladder can be
rewritten as
\begin{eqnarray}
H &=& \sum_{i=1}^{\frac{N}{2}}
\{ J_0(1- \delta) \mbox{\boldmath $S_{1,i}$} \cdot \mbox{\boldmath $S_{1,i+1}$}
+ J \mbox{\boldmath $S_{1,i}$} \cdot ( \mbox{\boldmath $S_{1,i+1}$} +
\mbox{\boldmath $S_{2,i+1}$} ) \nonumber\\
&+& J (\mbox{\boldmath $S_{1,i}$} +
\mbox{\boldmath $S_{2,i}$} ) \cdot \mbox{\boldmath $S_{2,i+1}$ } \}.
\end{eqnarray}
We define the dimer state $|\psi_{\rm dimer}\rangle$ as
$ |\psi_{\rm dimer}\rangle = |[1,2]_1\rangle|[1,2]_2\rangle \cdots
|[1,2]_{\frac{N}{2}}\rangle$, where
$|[1,2]_i\rangle$ denotes a singlet state constructed from $\mbox{\boldmath $S_{1,i}$}$ and
$\mbox{\boldmath $S_{2,i}$}$. Then $|\psi_{\rm dimer}\rangle$ can be shown to be the eigen
state of the Hamiltonian;
\begin{eqnarray}
H|\psi_{\rm dimer}\rangle = - \frac{1}{16}J_0(1- \delta)N|\psi_{\rm dimer}\rangle.
\end{eqnarray}
When $J_0 \ge 3J$, on the other hand, by applying the Rayleigh-Ritz variational
principle we can show that $\langle\psi_{\rm G}|H|\psi_{\rm G}\rangle \ge
- \frac{1}{16}J_0(1- \delta)N$. (Here $|\psi_{\rm G}\rangle$ is the ground state of the
Hamiltonian.) This proves that the dimer state $|\psi_{\rm dimer}\rangle$ is the exact ground
state of the system.

Since we notice that our Hamiltonian is equivalent to that of the \tone chain
\cite{tone1,tone2} with a next-nearest-neighbor (N-N-N) interaction $J$ as
illustrated in Fig.1(b), it is worthwhile to refer to something relating with this
chain. Here the \tone chain means an one-dimensional antiferromagnetic chain where
two $S= \frac{1}{2}$ and two $S=1$ spins are arranged alternatingly. This \tone
chain can be visualized if we cut off the bond $J$ ($J \to 0$) in Fig.1(b). This chain
without the N-N-N interaction has been studied extensively by Tonegawa {\it et al.}
\cite{tone1,tone2} and several results about its ground state had been obtained
already. In their works, it was shown that a phase transition between two phases of
the ground state of the chain is brought about by the introduction of the bond
alternation between the nearest-neighbor antiferromagnetic interactions. They found
also that this phase transition was a quantum phase transition between two different
kinds of singlet phases which are represented schematically in Fig.2(a) and 2(b) by
means of the valence-bond-solid picture. Our alternating spin ladder maps onto this
bond-alternating \tone chain if $J_0/J$ approaches to infinity
($J \to 0$ equivalently). The bond-alternation in the \tone chain, furthermore,
corresponds to the variation of $\delta$ in our model. In our spin ladder, therefore,
we reasonably expect
for these singlet phases of the ground state to appear, and can expect that the same
type of phase transition between these singlet phases will occur at a certain value
of $\delta$ in the parameter region where the ratio $J_0/J$ is large enough. After
the numerical calculation, it is found that the result is really the case.

\section{Numerical Results}

In order to clarify the ground state properties of the alternating spin ladder, we
make a numerical study of the system by means of the exact diagonalization method
exploiting the Lanczos technique \cite{titpack}. Because of the restriction of our
computational environments, the calculation has been carried out for the system
size of $N=8$, $12$, $16$, and $20$.

\subsection{Case of small $J_0/J$}

The system in this parameter region is regarded as a composite of two ferrimagnetic
chains interacting weakly each other via two kinds of the inter-chain interactions.
As stated in the previous section, for case of $\delta = 1$ and $-1$, we know that
the ground state becomes a quantum ferrimagnetic state and a singlet state,
respectively. Thus there should be a phase transition of the ground state at
$\delta = \delta_f$; the ground state is the quantum ferrimagnetic state for
$\delta > \delta_f$, and is the singlet state for $\delta < \delta_f$. This is
revealed by calculating the $\delta$-dependence of the energy gap between the ground
state and the first excited state. The result in the case of $J_0/J = 0.01$ is shown
in Fig.3. Since the energy gap in the singlet state is proportional to $\delta$, the
transition point $\delta_f$ can be determined easily by the least square method;
for $N=20$, $\delta_f = 0.21557$ which is only $0.5$ \% smaller than one
for $N = 16$, so that it would be close to the bulk limit. The phase transition
between the singlet state and the ferrimagnetic state is of the first order, because
the total spin of the system is found to change from zero to $\frac{N}{4}$ as $\delta$
goes from the region of $\delta < \delta_f$ into the region of $\delta > \delta_f$.

From additional calculations, we find the value of $\delta_f$ not to be sensitive to
the value of $J_0/J$ as far as $ J_0/J \simle 0.7 $;
that is, in this region of $J_0/J$, the
appearance of the ferrimagnetism depends scarcely on the ratio of the total amount of
the inter-chain interactions to the intra-chain interaction but does on the ratio of
one kind of inter-chain interaction to the other. As a result, the ferrimagnetism
disappears for $\delta < \delta_f$ however weak the inter-chain interaction is.

Because the total spin moment of each ferrimagnetic chain,
$\langle(\sum_{i=1}^{\frac{N}{2}}\mbox{\boldmath $S_{j,i}$})^2\rangle \; (j=1,2)$,
was confirmed to be close to $\frac{N}{8}$
in whole region of $\delta$, we should regard the energy gap in $\delta < \delta_f$ as
the singlet-triplet gap of a composite of two giant spins which come from respective
ferrimagnetic chains. Although such a state is far
from the dimer state which is described in \S 2, we would like to call the region
of these states as the dimer phase for convenience.

\subsection{Case of large $J_0/J$}

In the limit of $J \to 0$, our alternating spin ladder becomes identical to the \tone
chain which has been studied extensively by Tonegawa {\it et al.} \cite{tone1,tone2}.
Then, in the parameter region of large $J_0/J$, we can guess the phenomena in our
system by utilizing the knowledge of the phenomena in the \tone chain. This
confirmation, furthermore, provides a justification for numerical results obtained in
our system. One of the important results in the \tone chain is the existence of the
quantum phase transition between two types of singlet states, namely the dimer state
and VBS state which are usually depicted by the diagrams in Fig.2(a) and 2(b),
respectively \cite{tone1,tone2}. In the case of large $J_0/J$ in our system,
therefore, we expect reasonably that the same kind of the quantum phase transition
will also occur at an appropriate value of $\delta = \delta_c$. Tonegawa {\it et al.}
showed that the energy gap between the ground state and the first excited state of
their chain goes to zero at this transition point.

Keeping this knowledge in mind,
let us look up the calculated energy gap as functions of $\delta$ for
$J_0/J = 5.0$ with $N = 8$, $12$, and $16$ shown in Fig.4. We see that the transition
point between the singlet and the ferrimagnetic state moves toward right
($\delta_f = 0.87005$) comparing with the case of $J_0/J = 0.01$.
Among of all, however, we notice that there is a great difference between this result and that for
$J_0/J = 0.01$ in Fig.3; that is, the existence of a minimum in the energy gap. From
the knowledge of the \tone chain, we should consider this minimum signifies that the
same type of quantum phase transition occurs as has appeared in the chain, although our minimum value is
not zero. This non-zero value of the minimum in the energy gap is due to the
finiteness of the system size. We can show, in fact, that the minimum value in the
energy gap goes to zero as $N$ goes to infinity: We employ such a linear scaling
function in $1/N$ as
\begin{eqnarray}
E_{gap}(N)/J &=& E_{gap}/J + \frac{a}{N}
\end{eqnarray}
where $a$ is a constant of proportionality, $E_{gap}(N)$ denotes the minimum
value of the energy gap for the system size $N$
and $E_{gap}$ gives the energy gap in the bulk limit. By using data of $E_{gap}(N)$
($N=8$,$12$,and $16$) for several values of $J_0/J$, we extrapolate the energy gap
$E_{gap}(N)$ at $\delta = \delta_c$ to the balk limit ($1/N \to 0$). The results
are illustrated in Fig.5. In this way, we ensure that the thermodynamic limit of
the energy gap is really vanishing at $\delta = \delta_c$ and, at the same time,
that the value of $\delta_c$ approaches to $-0.129 \pm 0.001$ as $J_0/J$ increases.
This value of $\delta_c$ is consistent with the result calculated by
Tonegawa {\it et al.} in their chain. The property of the singlet phase in the region
of $\delta < \delta_c$ for each $J_0/J$ can be guessed as dimer like because the
region is continuously connected to the exact dimer line $J_0(1+ \delta ) = 2J$
for $J_0 \ge 3J$. The singlet phase in this region, therefore, should be call as
a ``dimer phase''.

In order to clarify the nature of a singlet phase whose region is sandwiched between
those of the dimer phase and the ferrimagnetic phase, we calculate three kinds of
inter-chain spin-spin correlation functions;
\begin{eqnarray}
&\omega&_{\frac{1}{2} \hyp \frac{1}{2}} =
\langle\mbox{\boldmath $S_{1,2i \hyp 1}$} \cdot
\mbox{\boldmath $S_{2,2i \hyp 1}$}\rangle,\\
&\omega&_{1 \hyp \frac{1}{2}} \, =
\langle\mbox{\boldmath $S_{1,2i}$} \cdot
\mbox{\boldmath $S_{2,2i \hyp 1}$}\rangle,\\
&\omega&_{1 \hyp 1} \,\, =
\langle\mbox{\boldmath $S_{1,2i}$} \cdot
\mbox{\boldmath $S_{2,2i}$}\rangle,
\end{eqnarray}
where $\omega$'s are inter-chain correlation functions between two spins whose
magnitudes are specified by their subscripts. In Fig.6, numerical results for these
correlation functions are exhibited as functions of $\delta$ for the case of
$J_0/J = 5.0$ with $N=16$. The results
shown in Fig.6, therefore, indicate that the singlet ground state in the region
$\delta < \delta_c$ is well described by the dimer state while one in the region
$\delta_c < \delta < \delta_f$ is done by the VBS state; in other words, the dimer
state and the VBS state provide a good trial wave function to the singlet ground
state in respective regions. We may, therefore, suitably call the singlet phase in
$\delta_c < \delta < \delta_f$ as the VBS phase. We can also see in Fig.6 that rapid
changes in correlation functions are occurring around $\delta = \delta_c$. From
these analyses, we may surely say that a phase transition between the dimer phase
and the VBS phase occurs at $\delta = \delta_c$.

\subsection{Case of $J_0/J = 1.5$}

First, in Fig.7, we show the $\delta$-dependence of the inter-chain spin-spin
correlation functions, $\omega_{\frac{1}{2} \hyp \frac{1}{2}}$, $\omega_{1 \hyp \frac{1}{2}}$
and $\omega_{1 \hyp 1}$ calculated for $J_0/J = 1.5$ and $N = 16$. It is noticed in this
figure that the region of the VBS phase becomes narrower than that for
$J_0/J = 5.0$. The curves in this figure, further, make us feel that the natures
of these singlet phases of the ground state for this $J_0/J$ value may not be
described suitably by such pictures as the dimer state and the VBS state. Nevertheless,
in the characteristic behavior of the curves in this figure, we find a kind of
similarity with that in the curves for $J_0/J = 5.0$ in Fig.6. So, even for the case
of $J_0/J = 1.5$, we use the terms such as the ``dimer phase'' and the
``VBS phase'' for convenience to call these singlet phases of the ground state
appearing in the region $\delta < \delta_f$.

In Fig.8, the $\delta$-dependence of the energy gap for $J_0/J = 1.5$ is shown. The
curves exhibit their minimum points expectedly at $\delta = \delta_c$, at which the
curves for the spin correlation functions in Fig.7 have been observed to make rapid
changes. This situation is somewhat similar to one in Fig.4 for large $J_0/J$. We
find, however, that the behavior of the energy gap in Fig.8 differs a bit from one
in Fig.4. Especially, there is a sharp minimum in the energy gap about
the boundary point between the
dimer phase and the VBS phase. Furthermore, we could not extrapolate the value of
the energy gap in its minimum to the bulk limit by such a function as given
in eq.(3.1). To make clear the size dependence of the energy gap, it will be necessary
that we calculate it in much larger system. From additional calculations, the VBS phase is
likely to disappear in the region where $J_0$ is comparable to $J$ (see also the phase
diagram in the next section). As a matter of course, the transition point between the ferrimagnetic
phase and the singlet phase, $\delta_f$, is observed in Fig.8, and we can read off
the value of $\delta_f$ to be 0.58310.

\section{Phase Diagram and Discussions}

The illustration of $\delta$-$J_0/J$ phase diagram would be helpful to understand
the properties of the ground state in our system and the relation between our system
and the other spin system. Based on the data shown in the previous section together
with additional ones, we make up the $\delta$-$J_0/J$ phase diagram for the ground state
properties, which is exhibited in Fig.9. We note that three phases are appearing
there; named by ``ferrimagnetic phase'',''dimer phase'', and ``VBS phase''.

Solid squares in the phase diagram express the points with the coordinates
($\delta_f$,$J_0/J$) which are determined by the values of $\delta_f$ for given
$J_0/J$; in the region $\delta > \delta_f$, the total spin moment is not zero so that
the ground states are ferrimagnetically ordered states while, for $\delta < \delta_f$,
the total spin moment becomes zero so that the ground state becomes non-magnetic.

In the region where the inter-chain interactions are weak, there exists only a phase
transition between the ferrimagnetic phase and the dimer phase. There, the ferrimagnetic
phase is stable only for $\delta > \delta_f$. From the phase diagram
we note also that, as far as $J_0/J \simle 0.7$, the value of $\delta_f$ is not
sensitive to
the ratio $J_0/J$; that is, which of the ferrimagnetic phase and the dimer phase to
be stabilized is determined only by the ratio between the two kinds of inter-chain
interactions regardless of their total amount. Although this result is obtained for
the alternating spin ladder with constituent spin $S=\frac{1}{2}$ and $1$, this
seems to be scarcely changed by other choice for the magnitudes of the constituent spins.
Then in the case of the Mn$^{II}$Cu$^{II}$ bimetallic chain compound, for example, the
ratio of one kind of inter-chain interaction (Mn-Cu) to the other (Mn-Mn and Cu-Cu),
is supposed to be an important parameter which determines whether the quantum
ferrimagnetic state is stable or not.

In the region of the phase diagram where $J_0/J$ is not small, we notice the appearance
of the VBS phase. This phase shares the singlet region with the dimer phase. These
phase regions are divided by an assembly of solid and open circles; a solid line
connecting the former circles and a broken line connecting the latter ones are drawn
for a guide to eyes. The solid circles are points which express $\delta_c$ for given
$J_0/J$ values at which the energy gaps have their minimum whose values are confirmed
to vanish in the bulk limit by extrapolation procedure with the scaling function
eq.(3.1).
Open circles, on the other hand, are points determined from the same minimum points but
their bulk limits could not be confirmed since the scaling function eq.(3.1)
does not fit with
our data; this may be due to that our system size is too small for the procedure to work.

We note also that the solid boundary line between the dimer phase and the VBS phase
approaches asymptotically to the line $\delta = -0.129$ as $J_0/J$ goes to infinity.
This value of $\delta$ corresponds to the ratio between the alternating bonds in the \tone
chain at which the chain has shown to make a quantum phase transition between the
``dimer'' ground state and the ``VBS'' ground state \cite{tone1,tone2}. Therefore, the result of
our calculations is surely consistent with the result revealed in the work of
Tonegawa {\it et al.} by means of a quantum Monte Calro calculation. We notice,
furthermore, that a dot-dashed line whose equation is expressed as $J_0(1+ \delta ) = 2J$
(with $J_0 > 3J$) is reasonably embedded in the dimer phase region; on this line, as
stated in \S 2, the dimer state becomes the exact ground state. The ground state may
be well described in the region close to this exact dimer line. Therefore, we can
understand that, as $J_0/J$ becomes large, such a picture of giant spins as
described in the case of small $J_0/J$ may break down eventually.

From all these considerations, we have derived a conclusion that, on the solid
portion of the boundary line, a quantum phase transition between the dimer and the
VBS ground state occurs as it does in the bond alternating \tone chain. This
conclusion may be claimed safely as far as $J_0/J > 2.5$. For $J_0/J \simle 2$,
it is not clear yet what kind of phase transition occurs on the
broken boundary line consisting of open circles. We, however, guess that the phase
transition on the broken line would be of the first order,
because we find a similar situation in the bond alternating Heisenberg model with
spin $S=1$ involving the next-nearest-neighbor interaction examined by others
\cite{firstorder1,firstorder2}. At present, our assertion on this point is only a conjecture. The
confirmation of it requires the calculations in much larger system. We think
that the method of the density matrix renomalization group may provide an
appropriate means to carry out this calculations.

The properties of our system
at finite temperatures, such as the specific heat and the magnetic susceptibility,
remain as the matters of further interest in connection with the possibility of the
ferromagnetic-antiferromagnetic crossover.

\section{Summary}

In order to investigate the ground state property of the alternating spin ladder, we carry
out the numerical study of the energy gap and the spin-spin correlation function by using the
exact diagonalization method. Here the spin ladder is composed of two alternating spin
chains of $S=\frac{1}{2}$ and $1$ having an antiferromagnetic intra-chain interaction.
Two kinds of antiferromagnetic interactions are introduced between the chains. Through the
competition among these interactions, it is revealed that the system provides an interesting
variety of phases in the ground state. By piling up the data of calculations, we make
up a phase diagram of the ground state in the parameter space $\delta$-$J_0/J$; $J_0/J$ is
the ratio of the amount of inter-chain interactions to the intra-chain interaction, and
$\delta$ is a measure of the ratio between two kinds of inter-chain interactions. In
the case of weak inter-chain interactions ($J_0/J \ll 1$), the ground state is shown to
undergo the first order transition between the ferrimagnetic phase and the singlet phase at
an appropriate value of $\delta = \delta_f$; $\delta_f = 0.21557$ for $J_0/J =0.01$. This
means that, however weak the inter-chain interactions is, there exists a region of $\delta$ where
the quantum ferrimagnetic state becomes unstable. In the region of large $J_0/J$, this first order
transition between the ferrimagnetic phase and the singlet phase is observed to occur at
much larger value of $\delta_f$ compared with the
case of small $J_0/J$, so that the ferrimagnetic state is seen to be destabilized more.
Furthermore, we see that the singlet phase is divided into the dimer phase and the VBS phase. The
transition between these phases is concluded to be of the second order as shown to be the case
in the \tone chain studied by Tonegawa {\it et al.} Even when $J_0/J$ becomes comparable to one,
it seems that the singlet phase region is also divided into a dimer like phase and a VBS like
one. We could not clarify the nature of the phase transition between them, at present. We
only speculate, however, that the phase transition between them may change its nature from
the second order transition to the first order one in this region.

\section*{Acknowledgements}
The authors would like to thank J.~Hosoi, K.~Sakai and Dr.~M.~Yamanaka
for helpful discussions and
encouragement. The authors appreciate very constructive comments by Dr.~Y.~Yokoya,
Dr.~Y.~Matsushita and Dr.~Y.~Suwa. Part of the numerical calculations in this work
have been performed
using the facilities of the Supercomputer Center, Institute for Solid State Physics,
University of Tokyo.

Fig.1

Schematic representation of the Hamiltonian (a) for the alternating spin ladder,
and (b) for the same Hamiltonian viewed as the \tone chain with the
next-nearest-neighbor-interaction $J$. Large and small circles represent
spins $S=1$ and $S=\frac{1}{2}$, respectively.

Fig.2

Ground state represented schematically by means of the VBS picture for the
\tone chain. (a) is for region of small $\delta$ and (b) is for region of large
$\delta$. The solid circles represent $S=\frac{1}{2}$ spin. The open circles denote
a triplet state made up of two $S=\frac{1}{2}$ spins in the circles. The solid lines
represent a singlet state of two $S=\frac{1}{2}$ spins on their ends.

Fig.3

$\delta$-dependence of the energy gap for $N=8$ (solid circle), $N=12$
(solid square), $N=16$ (solid triangle) and $N=20$ (open circle), in the case of
$J_0/J = 0.01$. The $\delta_f$ value indicated by an arrow in this figure is that
for the case $N=20$, which is read off as $0.21557$.

Fig.4

$\delta$-dependence of the energy gap for $N=8$ (solid circle), $N=12$
(solid square) and  $N=16$ (solid triangle), in the case of $J_0/J = 5.0$. The value
of $\delta_f$, 0.87005, indicated by an arrow is that for $N=16$, and the value of
$\delta_c$ indicated there is that for bulk limit ensured in Fig.5.

Fig.5

Extrapolation of the energy gap to its bulk limit with use of the scaling function eq.(3.1). 
In this figure, solid circles, solid squares, solid triangles, open circles,
open squares and open triangles represent $E_{gap}(N)/J$ for the set of parameters
$(J_0/J,\delta)$, $(5.0,-0.129)$, $(4.5,-0.129)$, $(4.0,-0.128)$,
$(3.5,-0.123)$, $(3.0,-0.116)$ and $(2.5,-0.104)$, respectively.

Fig.6

$\delta$-dependence of the inter-chain spin-spin correlation functions,
$\omega_{\frac{1}{2} \hyp \frac{1}{2}}$ (solid circle), $\omega_{1 \hyp \frac{1}{2}}$
(solid square) and $\omega_{1 \hyp 1}$ (solid triangle) for $J_0/J=5.0$ and
$N=16$. Broken lines in left hand side of this figure are indicating exact values
taken by $\omega_{1 \hyp \frac{1}{2}}$, $\omega_{\frac{1}{2} \hyp \frac{1}{2}}$ and
$\omega_{1 \hyp 1}$ in the dimer state from top to bottom. Broken lines in right hand
side are indicating exact values taken by $\omega_{\frac{1}{2} \hyp \frac{1}{2}}$,
$\omega_{1 \hyp \frac{1}{2}}$ and $\omega_{1 \hyp 1}$ in the VBS state from top to
bottom.

Fig.7

$\delta$-dependence of the inter-chain spin-spin correlation functions,
$\omega_{\frac{1}{2} \hyp \frac{1}{2}}$ (solid circle), $\omega_{1 \hyp \frac{1}{2}}$
(solid square) and $\omega_{1 \hyp 1}$ (solid triangle) for $J_0/J=1.5$ and
$N=16$. Broken lines in this figure has similar meanings to those in Fig.6.

Fig.8

$\delta$-dependence of the energy gap for $N=8$ (solid circle), $N=12$
(solid square) and $N=16$ (solid triangle), in the case of $J_0/J=1.5$. The indicated
value of $\delta_f$, 0.58310, in this figure is that for $N=16$.

Fig.9

Phase diagram for the ground state of the alternating spin ladder. Solid squares denote
the transition points between the ferrimagnetic phase and the singlet phase. Solid
circles in the singlet phase denote the transition points with vanishing an energy gap.
Open circles indicate the points at which the energy gap between the two singlet phases
take minimum values. Solid lines and a broken line are drawn for a guide to eyes.
On the dot-dashed line, the dimer state becomes the exact ground state of the
alternating spin ladder.


\begin{thebibliography}{99}
%[1]
\bibitem{haldane1}F.~D.~M.~Haldane:
Phys.~Lett. {\bf A93} (1983) 464.
%[2]
\bibitem{haldane2}F.~D.~M.~Haldane:
Phys.~Rev.~Lett. {\bf 50} (1983) 1153.
%[3]
\bibitem{nenp}J.~P.~Renard, M.~Verdaguer, L.~P.~Regnault, W.~A.~C.~Erkelens,
J.~Rossat-Mignod and W.~G.~Stirling:
Europhys.~Lett. {\bf 3} (1987) 945.
%[4]
\bibitem{gapmonte}M.~P.~Nightingale and H.~J.~Bl\"ote:
Phys.~Rev.~B {\bf 33} (1986) 659.
%[5]
\bibitem{matrix}A.~K.~Kolezhuk, H.~-J.~Mikeska and S.~Yamamoto:
Phys.~Rev.~B {\bf 55} (1997) R3336.
%[6]
\bibitem{low-temp}S.~Brehmer, H.~-J.~Mikeska and S.~Yamamoto:
J.~Phys.: Condens.~Matter {\bf 9} (1997) 3921.
%[7]
\bibitem{pati1}S.~K.~Pati, S.~Ramasesha and D.~Sen:
Phys.~Rev.~B {\bf 55} (1997) 8894.
%[8]
\bibitem{pati2}S.~K.~Pati, S.~Ramasesha and D.~Sen:
J.~Phys.: Condens.~Matter {\bf 9} (1997) 8707.
%[9]
\bibitem{mag1}Y.~Matsumoto, M.~Otani, H.~Miyagi and N.~Suzuki:
J.~Phys.~Soc.~Jpn. {\bf 66} (1997) 1844.
%[10]
\bibitem{ground}T.~Ono, T.~Nishimura, M.~Katsumura, T.~Morita and M.~Sugimoto:
J.~Phys.~Soc.~Jpn. {\bf 66} (1997) 2576.
%[11]
\bibitem{mag2}T.~Kuramoto:
J.~Phys.~Soc.~Jpn. {\bf 67} (1998) 1762.
%[12]
\bibitem{elementary}S.~Yamamoto, S.~Brehmer and H.~-J.~Mikeska:
Phys.~Rev.~B {\bf 57} (1998) 13610.
%[13]
\bibitem{thermo}S.~Yamamoto and T.~Fukui:
Phys.~Rev.~B {\bf 57} (1998) R14008.
%[14]
\bibitem{combi}S.~Yamamoto, T.~Fukui, K.~Maisinger and U.~Schollw\"ock:
J.~Phys.: Condens.~Matter {\bf 10} (1998) 11033.
%[15]
\bibitem{low-ene}S.~Yamamoto and T.~Sakai:
J.~Phys.~Soc.~Jpn. {\bf 67} (1998) 3711.
%[16]
\bibitem{mag3}S.~Yamamoto:
Phys.~Rev.~B {\bf 59} (1999) 1024.
%[17]
\bibitem{bimetallic1}O.~Kahn, Y.~Pei, M.~Verdaguer, J.~P.~Renard and J.~Sletten:
J.~Am.~Chem.~Soc. {\bf 110} (1988) 782.
%[18]
\bibitem{bimetallic2}P.~J.~van~Koningsbruggen, O.~Kahn, K.~Nakatani, Y.~Pei,
J.~P.~Renard, M.~Drillon and P.~Legoll:
Inorg.~Chem. {\bf 29} (1990) 3325.
%[19]
\bibitem{mag4}M.~Hagiwara, K.~Minami, Y.~Narumi, K.~Tatani and K.~Kindo:
J.~Phys.~Soc.~Jpn. {\bf 67} (1998) 2209.
%[20]
\bibitem{lieb}E.~Lieb and D.~C.~Mattis:
J.~Math.~Phys. {\bf 3} (1962) 749.
%[21]
\bibitem{shastry}B.~S.~Shastry and B.~Sutherland:
Phys.~Rev.~Lett. {\bf 47} (1981) 964.
%[22]
\bibitem{tone1}T.~Tonegawa, T.~Hikihara, T.~Nishino, M.~Kaburagi,
S.~Miyashita and H.~-J.~Mikeska:
J.~Magn.~Magn.~Mater. {\bf 177-181} (1998) 647.
%[23]
\bibitem{tone2}T.~Tonegawa, T.~Hikihara, M.~Kaburagi, T.~Nishino,
S.~Miyashita and H.~-J.~Mikeska:
J.~Phys.~Soc.~Jpn. {\bf 67} (1998) 1000.
%[24]
\bibitem{titpack}Our program was made on the basis of the TITPACK ver.2
but with a slight modification so as to be applicable to a mixed spin system.
%[25]
\bibitem{firstorder1}A.~Kolezhuk, R.~Roth and U.~Schollwo\"ck:
Phys.~Rev.~Lett. {\bf 77} (1996) 5142.
%[26]
\bibitem{firstorder2}A.~Kolezhuk, R.~Roth and U.~Schollwo\"ck:
Phys.~Rev.~B {\bf 55} (1997) 8928.
\end{thebibliography}
\end{document}